\documentclass[aps,prl,twocolumn,amsmath,amssymb,superscriptaddress,hidelinks]{revtex4-1}

\usepackage{bm}
\usepackage{verbatim} 

\usepackage{graphicx}
\usepackage[usenames,dvipsnames]{color}
\usepackage[colorlinks]{hyperref}
\hypersetup{
  colorlinks,
  citecolor=blue,
  linkcolor=blue,
  urlcolor=blue}

\newcommand{\be}{\begin{equation}}
\newcommand{\ee}{\end{equation}}
\newcommand{\bea}{\begin{eqnarray}}
\newcommand{\eea}{\end{eqnarray}}

\newcommand{\la}{\langle}
\newcommand{\ra}{\rangle}

\renewcommand{\vec}[1]{{\bf #1}}

\begin{document}

\renewcommand{\hbar}{\mathchar'26\mkern-9mu h}

\title{Anomalous skew-scattering nonlinear Hall effect and chiral  photocurrents in {\it PT}-symmetric antiferromagnets}

\author{Da Ma} 
\thanks{These authors contributed equally to this work.}
\affiliation{Division of Physics and Applied Physics, School of Physical and Mathematical Sciences, Nanyang Technological University, Singapore 637371}
\author{Arpit Arora} 
\thanks{These authors contributed equally to this work.}
\affiliation{Division of Physics and Applied Physics, School of Physical and Mathematical Sciences, Nanyang Technological University, Singapore 637371}
\author{Giovanni Vignale}
\affiliation{The Institute for Functional Intelligent Materials (I-FIM), National University of Singapore, 4 Science Drive 2, Singapore 117544}
\author{Justin C.W. Song}
\email{corresponding author: justinsong@ntu.edu.sg}
\affiliation{Division of Physics and Applied Physics, School of Physical and Mathematical Sciences, Nanyang Technological University, Singapore 637371}

\begin{abstract}
Berry curvature and skew-scattering play central roles in determining both the linear and nonlinear anomalous Hall effects. Yet in {\it PT}-symmetric antiferromagnetic metals, Hall effects from either intrinsic Berry curvature mediated anomalous velocity or the conventional skew-scattering process individually vanish. Here we reveal an unexpected nonlinear Hall effect that relies on both Berry curvature and skew-scattering working in cooperation. This anomalous skew-scattering nonlinear Hall effect (ASN) is {\it PT}-even and dominates the low-frequency nonlinear Hall effect for {\it PT}-symmetric antiferromagnetic metals. Surprisingly, we find that in addition to its Hall response, ASN produces helicity dependent photocurrents, in contrast to other known {\it PT}-even nonlinearities in metals which are helicity blind. This characteristic enables to isolate ASN and establishes new photocurrent tools to interrogate the antiferromagnetic order of {\it PT}-symmetric metals. 
\end{abstract}

\maketitle

Nonlinear response can be a powerful diagnostic of a material's intrinsic symmetries. A prime example is the nonlinear Hall effect that manifests in time-reversal invariant but inversion broken metals~\cite{Moore2010,Paco2015,Sodemann2015,XCXie2018,QiongMa2018,MakKinFai2019,XCXie2019,Konig2019,Nandy2019,Xiao2019,LiangFu2020,XCXie2021a,HyunsooYang2022}. Arising at second-order in an applied electric field, the nonlinear Hall effect is often attributed to quantum geometric properties of Bloch electrons such as the Berry curvature dipole (BCD)~\cite{Sodemann2015,QiongMa2018,MakKinFai2019} or skew-scattering processes \cite{XCXie2019,Konig2019,LiangFu2020,HyunsooYang2022}. Such nonlinearities can persist even in antiferromagnets (e.g., BCD nonlinear Hall effect~\cite{Tsymbal2020}) when both inversion ({\it P}) and time-reversal ({\it T}) symmetries are broken. However, an unusual situation occurs in 
antiferromagnets that respect the combination of {\it P} and {\it T} symmetries, i.e. {\it PT} symmetry~\cite{Godinho2018,DiXiao2021,ShengyuanYang2021}. 
Even though antiferromagnetism breaks {\it P} and {\it T} symmetries simultaneously, {\it PT} symmetry zeroes out net Berry flux and ensures that the BCD~\cite{ShengyuanYang2021} and conventional skew-scattering nonlinearities vanish~\cite{Yanase2020}. Can Berry curvature or skew-scattering play any role in Hall responses of {\it PT}-symmetric materials?

\begin{figure}
\centering
\includegraphics[width=\linewidth]{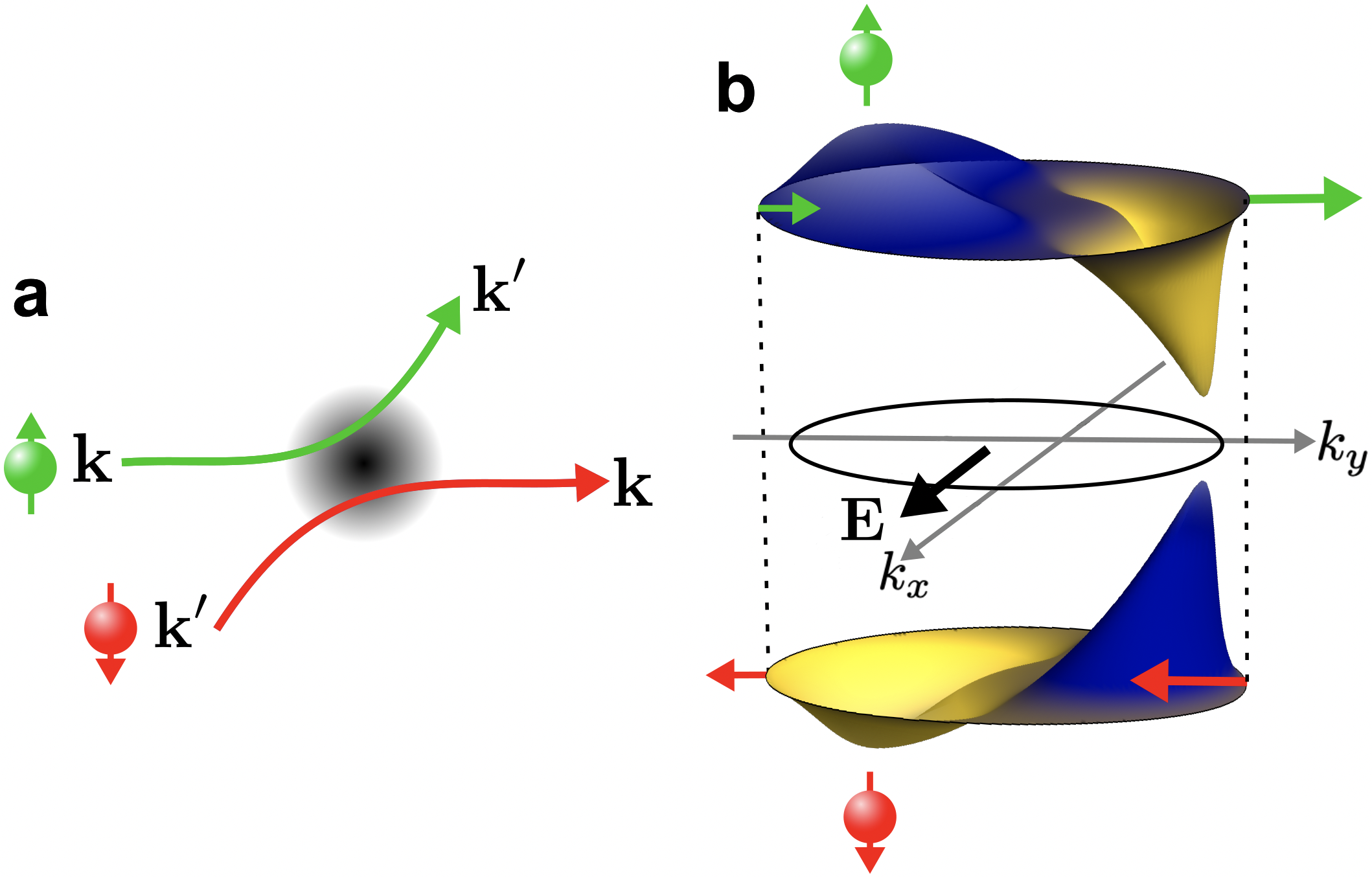}
\caption{Anomalous skew-scattering nonlinear Hall effect in {\it PT}-symmetric materials. (a) Scattering $\vec{k}\rightarrow\vec{k}'$ for up spin (green) shares the same rate as $\vec k' \rightarrow \vec k$ for down spin (red) in a {\it PT}-symmetric system. In particular, this {\it PT}-symmetry produces opposite skew-scattering contributions to the scattering rate for up and down spins respectively [see Eq.~(\ref{eq:rate})]. (b) Due to {\it PT} symmetry, the first-order skew-scattering driven deviations of the electronic distribution (from equilibrium, solid) have opposite signs for up and down spins; blue/yellow indicate sign of deviation. $PT$ symmetry also enforces opposite anomalous velocity for up (green horizontal arrows) and down (red horizontal arrows) spins. When this spin dependent anomalous velocity is combined with skew-scattering driven deviations of the electronic distribution, a non-vanishing nonlinear Hall effect manifests even in a $PT$-symmetric metal. }
\label{fig:Fermi-surface}
\end{figure}

Here we reveal a new paradigm for nonlinear transport where skew-scattering (extrinsic scattering) and Berry curvature (quantum geometric) {\it cooperate} to produce a second-order nonlinear Hall effect that persists in {\it PT}-symmetric materials. This anomalous skew-scattering nonlinear Hall effect (ASN) arises from combining a spin-dependent anomalous velocity and a skew-scattering spin-dependent distribution, Fig.~\ref{fig:Fermi-surface}b. ASN is {\it T}-odd, vanishing in {\it T}-symmetric materials; as such, it has been neglected. However, as we argue, ASN is {\it PT}-even, rendering {\it PT}-symmetric antiferromagnets a prime venue for its realization. 

Surprisingly, ASN also mediates a helicity dependent chiral photocurrent in the metallic limit peaking in the THz. This is striking since all other known intraband chiral photocurrents active in metals~\cite{Sodemann2015,Watanabe2021} 
vanish in {\it PT}-symmetric materials and are insensitive to magnetic ordering~\cite{Watanabe2021}. ASN, as we will see below, not only survives in {\it PT}-symmetric materials but is {\it T}-odd making it a useful new tool for accessing helicity dependent THz chiral photocurrents locked to magnetism.

Our work lies in the context of a recent surge of interest in second-order nonlinearities~\cite{DiXiao2021,ShengyuanYang2021,Yanase2020,Watanabe2021,Ahn-Nagaosa2020,Wang-QianNPJ2020,Zhang-BingHaiYan2019} in {\it PT}-symmetric antiferromagnets (e.g., CuMnAs \cite{Tang2016,Jungwirth2017}, MnBi$_2$Te$_4$~\cite{Otrokov2019,Su-YangXu2021}); such nonlinearities can be used to detect antiferromagnetic order, see e.g., Ref.~\cite{Godinho2018}. In the metallic/intraband limit, these have largely focussed on an intrinsic nonlinear Hall (INH) effect that arises from the Berry connection polarizability tensor~\cite{DiXiao2021,ShengyuanYang2021}. INH produces nonlinear Hall currents that are independent of the scattering time, $\tau$. In contrast, ASN is extrinsic and depends on $\tau$ at low frequencies. As a result, ASN is expected to dominate the nonlinear Hall effect in {\it PT}-symmetric antiferromagnets in the clean limit providing a much needed engineering strategy for boosting nonlinear Hall signals in {\it PT} antiferromagnets~\cite{Godinho2018}.

{\it PT partners and spin-dependent skew-scattering:} We begin by examining the effect {\it PT}-symmetry can have on the motion of electrons. As a simple illustration, consider the minimal Bloch hamiltonian $H^{(0)} (\vec k) = H_{\uparrow}^{(0)} (\vec k) + H_\downarrow^{(0)} (\vec k)$ where $s = \{ \uparrow, \downarrow\}$ are spins and $\vec k$ is the electron wavevector. {\it PT}-symmetry enforces double degeneracy and $(\vec P\vec T) H_{\uparrow}^{(0)} (\vec k) (\vec P\vec T)^{-1} = H_\downarrow^{(0)} (\vec k)$~\cite{Tang2016} yielding 
\be
\epsilon_{\uparrow} (\vec k) = \epsilon_{\downarrow}(\vec k) = \epsilon(\vec k), \quad \la  u_{\uparrow} (\vec k)  |  u_{\uparrow} (\vec k') \ra = \la u_{\downarrow} (\vec k')  |  u_{\downarrow} (\vec k)  \ra, 
\label{eq:PTsymmetry}
\ee
where $ |u_{s} (\vec k)\ra$ is a Bloch state of $H^{(0)} (\vec k)$ with a spin label $s$. For brevity of notation, we have omitted the band index. 
Eq.~(\ref{eq:PTsymmetry}) conveniently relates the properties of the {\it PT} partners $\uparrow$ and $\downarrow$. For example, $\uparrow$ and $\downarrow$ share the same group velocity $ \vec v (\vec k)= \partial_{\vec k} \epsilon (\vec k)/\hbar$, but possess opposite Berry curvature $\boldsymbol{\Omega}_{s} (\vec k) = i \la \nabla_{\vec k} u_{s} (\vec k) | \times | \nabla_{\vec k} u_{s} (\vec k) \ra $ signs. 

Eq.~(\ref{eq:PTsymmetry}) also constrains electronic scattering. In the presence of a scalar impurity potential $V$, the scattering rate in a single band is given by $W_{ \vec k \to \vec k'}^{s} = (2\pi/\hbar) | \la u_{s}(\vec k') | V | \psi_{s}(\vec k) \ra |^2 \delta [\epsilon_{s}(\vec k) - \epsilon_{s}(\vec k')]$~\cite{Nagaosa2010} that captures skew-scattering processes that occur beyond the Born approximation. Here $ |\psi_{s}(\vec k) \ra$ is an eigenstate of the full hamiltonian $H^{(0)} (\vec k) + V$ and can be expanded order-by-order using the self-consistency relation: 
$ |\psi_{s}(\vec k) \ra =  |u_{s}(\vec k) \ra +  [\epsilon_{s}(\vec k) - H_0(\vec k) + i\eta]^{-1} V |\psi_{s}(\vec k) \ra$~\cite{Nagaosa2010}. For scalar impurities and elastic scattering, we find (see {\bf SI} \cite{SeeSI}), 
\be
W_{\vec k \to \vec k'}^{\uparrow} = W_{\vec k' \to \vec k}^{\downarrow}, \quad w^{({\rm S}, {\rm A})}_{\uparrow, \vec k, \vec k'} = \pm w^{({\rm S}, {\rm A})}_{\downarrow, \vec k, \vec k'}, 
\label{eq:rate}
\ee
where $w^{({\rm S}, {\rm A})}_{s, \vec k, \vec k'} = [W_{\vec k' \to \vec k}^{s} \pm W_{\vec k \to \vec k'}^{s}]/2$ are the symmetric and skew (antisymmetric) scattering contributions to the total scattering rate respectively. Crucially, the scattering process $\uparrow, \vec k \to \uparrow, \vec k'$ is the {\it PT} partner of $\downarrow, \vec k' \to \downarrow, \vec k$ and have the same rate (Fig.~\ref{fig:Fermi-surface}a). As a result, the $\uparrow$, $\downarrow$ have {\it opposite} skew-scattering contributions. This conclusion persists for any {\it PT} symmetric scattering potential.

Eq.~(\ref{eq:rate}) applies order-by-order in $V$, and can be obtained by employing Eq.~(\ref{eq:PTsymmetry}) to the scattering rate. For an intuitive illustration of the origins of Eq.~(\ref{eq:rate}), we examine the familiar third order in $V$ expression for the skew-scattering rate~\cite{Luttinger1958,Sinitsyn2006,Nagaosa2010}
\be
w^{{\rm A}}_{s, \vec k, \vec k'} = \frac{4\pi^2 n_{i} V_0^3}{\hbar} \sum_{\vec k''} \delta_{\vec k, \vec k'}^{(\varepsilon)} \delta_{\vec k, \vec k''}^{(\varepsilon)} {\rm Im} \{ L_{s} (\vec k, \vec k'', \vec k')\},
\label{eq:skewthird}
\ee
where $V_0$ is the impurity strength, $n_i$ the impurity density,  $\delta_{\vec k, \vec k'}^{(\varepsilon)} = \delta [\epsilon(\vec k) - \epsilon(\vec k')]$, and $L_{s} (\vec k, \vec k'', \vec k') =  \la u_{s}(\vec k)| u_{s}(\vec k'') \ra \la u_{s}(\vec k'') | u_{s}( \vec k')\ra \la u_{s}(\vec k')| u_{s}(\vec k) \ra $ is the Wilson loop associated with the Pancharatnam-Berry phase of the skew-scattering process~\cite{Sinitsyn2006}. Directly applying Eq.~(\ref{eq:PTsymmetry}) to Eq.~(\ref{eq:skewthird}) yields a sign changing $w^A$ in Eq.~(\ref{eq:rate}). The inclusion of both {\it PT} partners (in our case, spin) is essential since applying the same reasoning to a spinless system produces a vanishing $w^A$ (e.g., Ref.~\cite{Yanase2020} computed a vanishing $w^A$ to $V^4$ in a spinless system).

Eq.~(\ref{eq:PTsymmetry}) and ~(\ref{eq:rate}) have a profound impact on transport behavior of {\it PT}-symmetric materials (e.g., {\it PT}-symmetric antiferromagnets). Because $\boldsymbol{\Omega}_{\uparrow} (\vec k) = - \boldsymbol{\Omega}_{\downarrow} (\vec k)$, the net Berry flux and the net Berry curvature dipole (BCD) vanish thereby zeroing out the intrinsic linear anomalous Hall as well as the BCD nonlinear Hall effect. Similarly, the changes to the distribution function due to skew-scattering in Eq.~(\ref{eq:rate}) are opposite for $\uparrow$ and $\downarrow$ (see Fig.~\ref{fig:Fermi-surface}b and detailed discussion below); when combined with $ \vec v (\vec k)= \partial_{\vec k} \epsilon (\vec k)/\hbar$, the conventional skew-scattering anomalous Hall effect at both linear and second order vanishes under {\it PT}-symmetry. 

\begin{table}
\begin{center}
\begin{tabular}{l c c c}
\hline\hline
Nonlinear Hall effects & $T$ & $PT$ & references \\
\hline 
Berry curvature dipole (BCD)                    & $+$ & $-$         & \cite{Sodemann2015,QiongMa2018,MakKinFai2019,Tsymbal2020} \\
Intrinsic (INH)                        & $-$        & $+$ & \cite{DiXiao2021,ShengyuanYang2021} \\
Conventional skew-scattering           & $+$ & $-$         & \cite{XCXie2019,Konig2019,LiangFu2020} \\
Anomalous skew-scattering (ASN) *              & $-$        & $+$ & this work \\  
\hline \hline
\end{tabular}
\end{center}
\caption{Symmetry of intraband nonlinear Hall responses. $+$ indicates the response is even (i.e. allowed by symmetry), $-$ means it is odd (i.e. forbidden by symmetry). Starred nonlinear Hall susceptibility is the new {\it PT}-even response discussed in this work in Eq.~(\ref{eq:ASN}) for the ASN, see also \textbf{SI}~\cite{SeeSI}.}
\label{Table:Classification}
\end{table}

{\it Anomalous skew-scattering nonlinear Hall effect:} However, when both Berry curvature mediated anomalous velocity ({\it PT}-odd) as well as the changes to the distribution function driven by skew-scattering ({\it PT}-odd) combine, a non-vanishing second-order ASN Hall effect ({\it PT}-even) can be produced (Fig.~\ref{fig:Fermi-surface}b). To see this in a systematic fashion, we analyze the net charge current 
\be
\vec j (t) =  - e \sum_{\vec k, s} \big(\vec v (\vec k) + e \boldsymbol{\mathcal{E}} (t)/\hbar \times  \bar{\boldsymbol{\Omega}}_{s} (\vec k) \big) f_{s} (\vec k,t), 
\label{eq:current} 
\ee
where $-e<0$ is the carrier charge, $\boldsymbol{\mathcal{E}} (t)$ is a time-varying uniform electric field, $f_{s} (\vec k,t)$ is the distribution function, and $\bar{\boldsymbol{\Omega}}_s (\vec k)$ is the modified Berry curvature that includes both intrinsic Bloch band Berry curvature [$\boldsymbol{\Omega_{s}} (\vec k)$] as well as field-induced corrections~\cite{QianNiu2014,DiXiao2021,ShengyuanYang2021} 
\be
\bar{\boldsymbol{\Omega}}_{s} (\vec k) = \boldsymbol{\Omega}_{s} (\vec k) + \nabla_{\vec k} \times \boldsymbol{\mathcal{G}} (\vec k)\boldsymbol{\mathcal{E}} (t). 
\label{eq:berry}
\ee
Here $\boldsymbol{\mathcal{G}} (\vec k)$ is the Berry-connection polarizability tensor in the metallic band of interest~\cite{QianNiu2014,DiXiao2021,ShengyuanYang2021}. For band $n$, $[\boldsymbol{\mathcal{G}}]_{ab} (\vec k)  = 2 e {\rm Re} \{\sum_{n'\neq n} A_a^{nn'} (\vec k) A_b^{n'n} (\vec k)/ [\epsilon_{n}(\vec k) - \epsilon_{n'}(\vec k)]\}$, with $ A_b^{nn'} = \la u_{n} (\vec k) | i\partial_{k_b} u_{n'}(\vec k)  \ra$. Here, $a,b$ denotes Cartesian coordinates and $\boldsymbol{\mathcal{G}} (\vec k)$ is even under {\it PT}. 

For clarity, we concentrate on the intraband limit and focus on scalar impurities that preserve {\it PT} symmetry and conserve spin. The distribution function in Eq.~(\ref{eq:current}) can be directly computed via a spatially uniform kinetic equation, $\partial_t f_s(\vec k,t) - e \boldsymbol{\mathcal{E}} (t) \cdot \partial_\vec k f_s(\vec k, t)/\hbar = I \{ f_s (\vec k,t) \} $ where \cite{Kohn-Luttinger1957,Sinitsyn2007,Nagaosa2010}
\be
I \{ f_s (\vec k,t) \} = \sum_{\vec k'} \big [W^s_{\vec k' \to \vec k } f_s (\vec k',t) - W^s_{\vec k \to \vec k'} f_s (\vec k,t)\big]  
\label{eq:KE}
\ee
describes the spin-dependent collision integral. 

The distribution function can be solved in the standard perturbative fashion: in powers of $\boldsymbol{\mathcal{E}}$ and for weak skew-scattering by using the relaxation time approximation, see {\bf SI} for a detailed derivation~\cite{SeeSI}. As such, we expand $f_s(\vec k,t)$ as 
\be
f_{s} (\vec k,t) = f_{0}(\vec k) + \sum _{\ell, m} f_{\ell ,s}^{(m)} (\vec k,t), 
\label{eq:generalf}
\ee
where the second term captures the deviation of the distribution function from the equilibrium distribution function, $f_{0}(\vec k)$. Here subscript $\ell = 1, 2, \cdots $ denote order in $\boldsymbol{\mathcal{E}}$ and the superscript $m = 0, 1, \cdots$ denote its dependence on skew-scattering rate;
$m=0$ captures the purely symmetric part independent of skew-scattering. 

Note $f_{0}(\vec k)$ is the same for both $\uparrow$ and $\downarrow$ due to {\it PT} symmetry. Similarly, even as $m=0$ contributions depend on the (transport) relaxation time: $\left( \tau^s \right)^{-1} = \langle \sum_{\vec k'} w^{S}_{s, \vec k', \vec k} \left( 1 - \cos\theta_{\vec{v} \vec{v}' } \right) \rangle$, {\it PT} symmetry in Eq.~(\ref{eq:rate}) ensure $f_{\ell, s}^{(0)} (\vec k, t)$ are the same for $\uparrow$ and $\downarrow$ since $\tau^\uparrow = \tau^\downarrow = \tau$. Here $\theta_{\vec{v} \vec{v}' }$ is the angle between $\vec{v}(\vec{k})$ and $\vec{v}(\vec{k}')$, and $\langle \cdots \rangle$ indicates an average over an energy contour. In contrast, skew-scattering  ($m=1$) contributions to the distribution function $f_{\ell,s}^{(1)} (\vec k,t)$ have opposite signs for opposite spins, see Fig.~\ref{fig:Fermi-surface}b: a property key to ASN.

Writing $\boldsymbol{\mathcal{E}}(t) = \vec E e^{i\omega t} + c.c.$ and substituting the distribution functions into Eq.~(\ref{eq:current}) enables to directly discern the nonlinear Hall responses. Amongst the possible second-order nonlinear Hall responses obtained (see Table~\ref{Table:Classification}), two are {\it PT}-even; the rest are odd. The first {\it PT}-even response is the intrinsic nonlinear Hall (INH) effect~\cite{QianNiu2014,DiXiao2021,ShengyuanYang2021} obtained by combining the second term in Eq.~(\ref{eq:berry}) with $f_0(\vec k)$. This yields an INH current 
$[j^{\rm INH}]_a (t) = {\rm Re}( j_a^0 + j_a^{2\omega} e^{i2\omega t})$ with $ j_{a}^0  = \chi_{abc}^{\rm INH}  [E_b]^* E_c$ and $ j_{a}^{2\omega}  = \chi_{abc}^{\rm INH}  E_b E_c$, 
where $\chi^{\rm INH}_{abc}$ \cite{QianNiu2014,DiXiao2021,ShengyuanYang2021} depends only on band geometric quantities. $\chi^{\rm INH}_{abc}$ is independent of $\tau$ and insensitive to $\omega$ in the semiclassical limit. 

The second {\it PT}-even nonlinear Hall response, ASN, is the main result of our work. 
This nonlinear Hall effect arises from combining $\boldsymbol{\Omega}_s(\vec k) \times \boldsymbol{\mathcal{E}}(t)$ with the skew distribution function $f_{1,s}^{(1)} (\vec k,t)$. This produces a nonlinear Hall response: $[j^{\rm ASN}]_a (t) = {\rm Re}( j_a^0 + j_a^{2\omega} e^{i2\omega t})$ with $j_a^0 = \chi_{abc}^{\rm ASN} [E_b]^* E_c$ and $j_{a}^{2\omega}  = \chi_{abc}^{\rm ASN}  E_b E_c$ with 
\be
\chi_{abc}^{{\rm ASN}} = 2 \frac{e^3\varepsilon_{adb}}{\hbar^2}\sum_{\vec k, \vec k', s} \Omega^s_d(\vec k) \tilde \tau_{\omega}^{2} w^A_{s,\vec k,\vec k'} \left[\frac{\partial f_0 (\vec k')}{\partial \vec k'}\right]_c,
\label{eq:ASN}
\ee 
where $\Omega^s_d (\vec k)$ denotes the $d$ component of $\boldsymbol{\Omega}_s (\vec k)$, $\tilde \tau_\omega = \tau/(1+i\omega \tau)$ and $\varepsilon_{adb}$ is the Levi-Civita symbol. Since $\Omega^s_d $ and $w^A_{s}$ are both odd under ${\it PT}$, their product is even producing a finite extrinsic nonlinear Hall effect. Importantly, $\chi_{abc}^{{\rm ASN}}$ scales as $\tau^2 w^A$ for $\omega \tau \ll 1$. As a result, $\chi_{abc}^{{\rm ASN}}$ is expected to dominate the nonlinear Hall response in {\it clean} systems. At finite $\omega$,  $\chi_{abc}^{{\rm ASN}}$ displays a characteristic $\omega$ dependence varying rapidly on the scale $1/\tau$ (see below); this $\omega$ dependence distinguishes it from both the $\omega$ insensitive $\chi^{\rm INH}$ as well as interband effects that have characteristic $\omega$ dependence on the scale of interband transition energy $\epsilon_n - \epsilon_m$. 

{\it Symmetry, scattering, and chiral photocurrents:} ASN has several striking attributes. Due to its Berry curvature roots, $\chi_{abc}^{{\rm ASN}}$ is antisymmetric in its first two indices yielding a nonlinear Hall effect~\cite{Souza2022} always transverse to the applied electric field. This antisymmetric nature imposes additional point-group symmetry constraints as compared to conventional skew-scattering nonlinearities~\cite{LiangFu2020,XCXie2019,HyunsooYang2022}. For example, in 2D, antisymmetric nonlinear $\chi_{abc}$ requires broken rotational symmetry~\cite{ShengyuanYang2021,Sodemann2015}. 

ASN's antisymmetric behavior contrasts with that of another {\it PT}-even nonlinear response that arises from combining $\vec v(\vec k)$ with $f_{2,s}^{(0)} (\vec k,t)$ \cite{Yanase2020,DiXiao2021} to produce a classical nonlinearity, $\chi_{abc}^{\rm Drude}$. Importantly, $\chi_{abc}^{\rm Drude}$ has a susceptibilty that is completely {\it symmetric} when its indices are permuted yielding a response that need not always be transverse as required of Hall type responses~\cite{Souza2022}. Experimentally, this fully symmetric nonlinear Drude response can be weeded out via interchanging the directions of driving field and response: symmetric $\chi_{abc}$ is even under exchange, whereas nonlinear Hall responses are odd.

Perhaps most striking is how ASN produces a helicity-dependent chiral photocurrent: 
$
[j^{\circlearrowleft}]_a = \frac{i}{2} {\rm Im}[\chi_{abc}](E_b^* E_c - E_b E_c^*). 
$
ASN chiral photocurrent arises from its part quantum geometric and part skew-scattering origins. First, since ASN depends on skew scattering $\chi_{abc}^{\rm ASN}$ possesses {\it both} real and imaginary components arising from the complex valued $\tilde{\tau}_{\omega}^2$ in Eq.~(\ref{eq:ASN}). Second, because ASN proceeds from the anomalous velocity $\boldsymbol{\Omega}\times \vec{E}$, its susceptibility is asymmetric allowing for a non-zero $\vec j^{\circlearrowleft}$ after both $b$ and $c$ indices are summed.

Importantly, ASN's combination of geometric nature and scattering processes is essential. For instance, even as symmetric scattering alone enables a nonlinear Drude conductivity $\chi_{abc}^{\rm Drude}$~\cite{Yanase2020,DiXiao2021} that has an imaginary component, it nevertheless is completely symmetric under any interchange of indices yielding a zero $\vec j^{\circlearrowleft}$. Similarly, while $\chi^{\rm INH}_{abc}$ is also asymmetric, it nevertheless is purely real producing helicity blind photocurrents. 
As a result, to our knowledge, $\chi_{abc}^{{\rm ASN}}$ is the only {\it intraband} nonlinearity that produces a helicity dependent chiral photocurrent in {\it PT}-symmetric antiferromagnets, see below for a discussion of interband effects.

{\it ASN in two-dimensional PT-even antiferromagnets:} To illustrate ASN, we adopt a minimal spinful model where both {\it P} and {\it T} symmetries are simultaneously broken, but composite {\it PT} symmetry is preserved. {\it PT} enforced doubly degenerate bands can be modelled by spinful massive Dirac fermions~\cite{Tang2016} 
\begin{equation}
\label{eq:effectivemodel}
H = \hbar v k_{x} \sigma_{x} + \hbar v k_{y} \sigma_{y}  + \Delta \sigma_{z} {s}_{z} + \hbar \beta v k_{y}, 
\end{equation}
where the Pauli matrices $\sigma$  and $s$ describe orbital and spin degrees of freedom respectively. Here $\Delta$ opens up a gap, $v$ is a velocity, and $\beta$ tilts the Dirac cone. 
The tilt term breaks rotational symmetries but preserves {\it PT}. Models like Eq.~(\ref{eq:effectivemodel}) were recently used to successfully capture the behavior of {\it PT}-symmetric antiferromagnets~\cite{Kaplan2020,ShengyuanYang2021}. Spinful  
Dirac fermions can be found in a variety of materials e.g., CuMnAs~\cite{Tang2016,Jungwirth2017}, even-layer MnBi$_2$Te$_4$~\cite{Otrokov2019,Su-YangXu2021}, as well as the antiferromagnet nodal line metal MnPd$_{2}$~\cite{Tsymbal2019}. While we concentrate on a simple model in Eq.~(\ref{eq:effectivemodel}) to illustrate ASN, our conclusions persist for more complex situations, e.g., an effective model of Dirac fermions in tetragonal CuMnAs~\cite{Jungwirth2017}, see {\bf SI} \cite{SeeSI}.

\begin{figure}
\centering
\includegraphics[width=\linewidth]{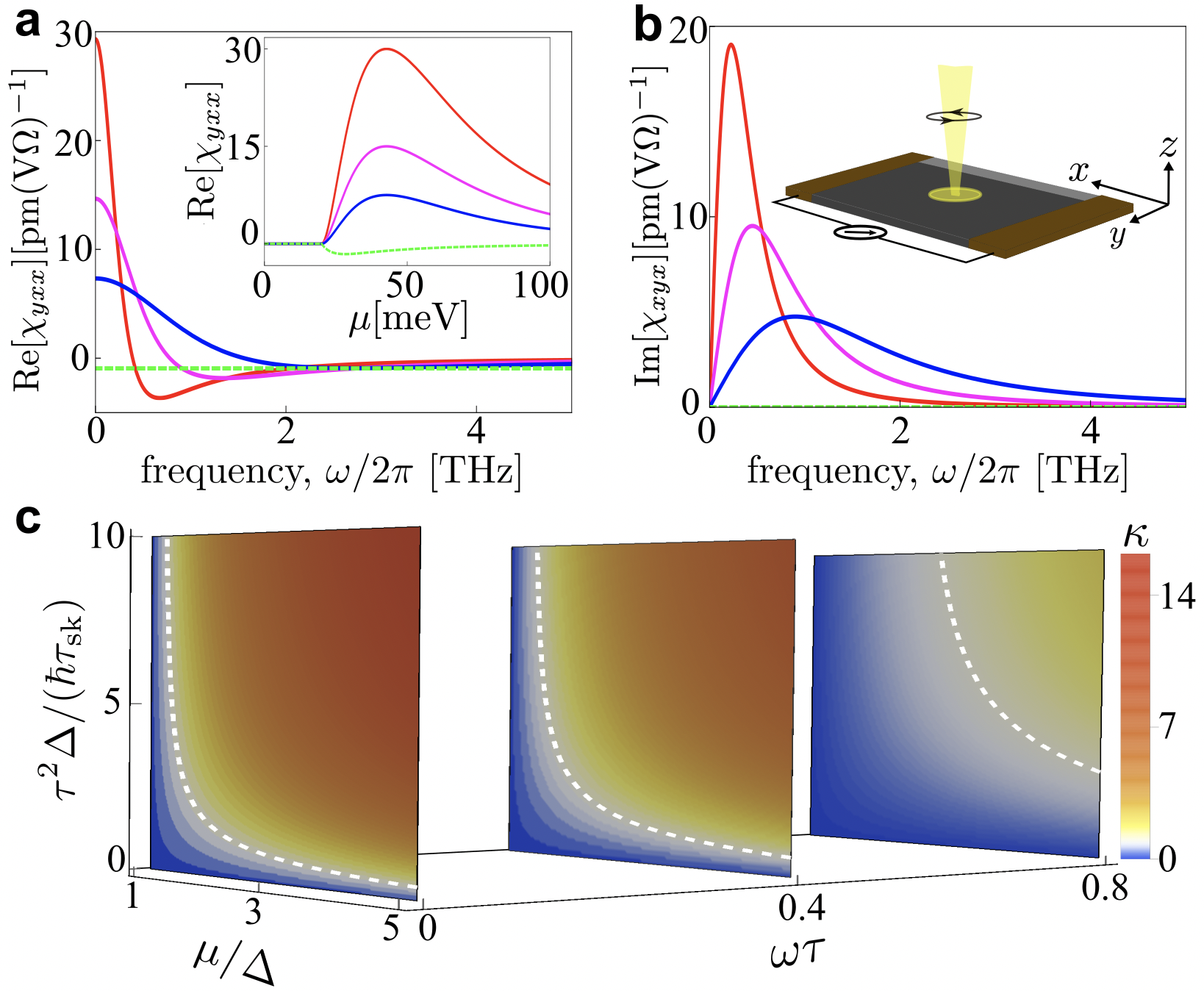}
\caption{ASN in a {\it PT}-symmetric antiferromagnetic metal. (a) Real part of the ASN susceptibility $\chi_{yxx}^{{\rm ASN}} = - \chi_{xyx}^{{\rm ASN}}$ (solid lines) at different driving frequencies for Hamiltonian in Eq.~(\ref{eq:effectivemodel}) dominates over $\chi^{\rm INH}$ (green dashed). (inset) ${\rm Re} \chi_{yxx}^{{\rm ASN}} (\omega \to 0)$ displays a peak away from the band bottom. 
(b) Imaginary part of ASN susceptibility can mediate a helicity dependent photocurrent (see text) and displays a maximal response at an intermediate frequency. (inset) ASN susceptibility can be isolated by probing the intraband helicity dependent photocurrent response. In both panels, red, magenta, and blue denote density of impurities set as $n_{\mathrm{i}} = 1, 2, 4 \ (\times 10^{9} \ \mathrm{cm}^{-2})$ that correspond to $\tau \approx 400, 200, 100 \ \mathrm{fs}$ respectively. 
(c) The ratio of magnitude of ASN to INH $\kappa$, depends on dimensionless parameters see text. The dashed white curve corresponds to $\kappa =1$. Parameters: for (a) and (b), $\Delta = 20 \ \mathrm{meV}$, $\mu = 50 \, {\rm meV}$ and $V_{0} = 6.2 \times 10^{-13} \ \mathrm{cm}^{2} \mathrm{eV}$; for all three panels,  $v = 10^{6} \ \mathrm{m/s}$, $\beta=0.1$. } 
\label{fig:Responses}
\end{figure}

Broken rotational symmetry [tilt $\beta$ in Eq.~(\ref{eq:effectivemodel})] is essential in enabling a non-vanishing $\chi^{\rm ASN}$ to develop in 2D (see discussion above); indeed, $\beta\neq 0$ means that anomalous velocities accrued at opposite ends of the Fermi surface (see Fig.~\ref{fig:Fermi-surface}b) do not cancel. Nevertheless, the {\it PT} and mirror symmetries of Eq.~(\ref{eq:effectivemodel}) still constrain $\chi^{\rm ASN}$: its only non-vanishing components are $\chi_{yxx}^{{\rm ASN}} = - \chi_{xyx}^{{\rm ASN}}$. To demonstrate ASN, we plot the second-order nonlinear susceptibility in Eq.~(\ref{eq:ASN}) for chemical potential in the conduction band of Eq.~(\ref{eq:effectivemodel}) in Fig.~\ref{fig:Responses} to leading order in $\beta$. In so doing, we have used short range impurities ${V}(\bm{r}) = V_{0} \sum_{j} \delta \left( \vec{r} - \vec{r}_{j} \right)$, with strength $V_0$ and impurity concentration $n_{\rm i}$, see caption for parameter values. 

In the low frequency limit $\omega \tau \ll 1$, ASN scales as $\tau^{2}w^{A}$ and grows with increasing $\tau$, while INH, as an intrinsic response, is independent of $\tau$.  As a result, we find that ASN dominates the nonlinear Hall effect in clean {\it PT}-symmetric materials (see Fig.~\ref{fig:Responses}a) with a peak like structure as a function of chemical potential [Fig.~\ref{fig:Responses}a(inset)]. It displays a sensitive dependence on frequency: its real part changes sign at $\omega=1/\tau$. This non-monotonic dependence, as well as its $\tau$ scaling can be used as a simple diagnostic of its manifestation. In Fig.~\ref{fig:Responses}c, we find that ASN dominates over wide swathes of the parameter space; the dimensionless ratio $\kappa = \vert \mathrm{Re}\chi_{yxx}^{{\rm ASN}} \vert/ \vert \mathrm{Re}\chi_{yxx}^{{\rm INH}} \vert $ is controlled by three dimensionless quantities: the dimensionless Fermi level $\mu/\Delta$, dimensionless frequency $\omega \tau$, and $\tau^{2}\Delta/(\tau_{\mathrm{sk}}\hbar) = \tau/\tau_{\rm sk} \times (\tau \Delta/\hbar)$ capturing the product of the characteristic skew scattering strength and a characteristic Compton-like scale that describes the effectiveness of the Berry curvature. We have used dimensionless quantities. Here the characteristic skew-scattering strength $\tau/\tau_{\rm sk} = 2\pi V_0 \nu (\Delta)$ where $\nu (\epsilon)$ is the density of states. The region of $\kappa>1$ is largest at small frequencies but still covers sizeable areas even for larger frequencies. 

Arising from the complex valued $\tilde{\tau}_\omega^2$ in Eq.~(\ref{eq:ASN}), ${\rm Im}\chi_{abc}^{\rm ASN}$ peaks when $\omega \sim 1/\tau$ [see Fig.~\ref{fig:Responses}b]. Strikingly, peak ${\rm Im}\chi_{abc}^{\rm ASN}$ is on par with ${\rm Re}\chi_{abc}^{\rm ASN}$ maximum. For typical $\tau$, ${\rm Im}\chi_{abc}^{\rm ASN}$ produces a chiral photocurrent peaked in the THz regime. Interestingly, in the {\it interband} regime at larger frequencies, other chiral photocurrents in {\it PT} antiferromagnets can also arise~\cite{Ahn-Nagaosa2020,Bhalla2023,Watanabe2021}. In particular, interband transitions can activate a circularly polarized light induced gyration current~\cite{Watanabe2021} that is {\it PT}-even but {\it T}-odd (also known as the circular shift photocurrent~\cite{Ahn-Nagaosa2020}); the gyration current corresponds to an imaginary nonlinear susceptibility. Importantly, gyration currents possess an $\omega$ dependence that tracks interband transitions (with characteristic scales $\omega \sim 2\Delta/\hbar$). This enables to distinguish from that of ASN chiral photocurrent that features characteristic frequency dependence in the {\it intraband} regime ($ \omega \sim 1/\tau$), see Fig.~\ref{fig:Responses}b.

ASN arises from the cooperative action of skew-scattering and Berry curvature; both are individually {\it PT}-odd, but when combined, produce a {\it PT}-even nonlinear Hall effect that can dominate over the currently known intrinsic mechanisms~\cite{DiXiao2021,ShengyuanYang2021} in the clean limit. This provides an engineering strategy (i.e. making the metal cleaner) for boosting the nonlinear Hall signals in {\it PT} antiferromagnets for more sensitive detection. Indeed, we estimate ASN provides sizeable nonlinear susceptibilities (see Fig.~\ref{fig:Responses} for values) on par with those recently measured in other nonlinear materials~\cite{MakKinFai2019}. Perhaps most striking is ASN's ability to mediate a helicity dependent photocurrent response enabling it to be directly isolated using circularly polarized drive fields. This can provide new tools for accessing a new type of quantum geometric opto-electronics~\cite{Shi2022,ma2023photocurrent} and pronounced nonlinearities in antiferromagnets. 

{\it Acknowledgements:} We acknowledge useful conversations with Mark Rudner and Roberto Raimondi. This work was supported by Singapore MOE Academic Research Fund Tier 3 Grant MOE2018-T3-1-002 and a Nanyang Technological University start-up grant (NTU- SUG).

\bibliographystyle{apsrev4-1}
\bibliography{Bib}


\onecolumngrid
\newpage
\pagebreak
\widetext
\begin{center}
\textbf{\large Supplementary Information for ``Anomalous skew-scattering nonlinear Hall effect and chiral photocurrents in {\it PT}-symmetric antiferromagnets''}
\end{center}

\setcounter{equation}{0}
\setcounter{table}{0}
\setcounter{figure}{0}
\makeatletter 
\renewcommand{\thefigure}{S\arabic{figure}}
\renewcommand{\theequation}{S\arabic{equation}}
\renewcommand{\thetable}{S\arabic{table}}
\renewcommand{\bibnumfmt}[1]{[S#1]}
\setcounter{page}{1}


\subsection{Scattering rate for $PT$ partners}

In this section we examine the scattering rate of {\it PT} partners. As in the main text, we focus on a spin block diagonal material described by $s=\{\uparrow,\downarrow\}$ and analyze spin preserving processes such as scattering from scalar impurity potential $\hat{V}$; here $\bar{s}$ is the {\it PT} partner of $s$. We proceed by writing the scattering rate for the process $(\vec{k}',s)$ to $(\vec{k},s)$ as~\cite{Nagaosa2010}
\begin{equation}
\label{eq:gold}
W^s_{\vec{k}'\rightarrow\vec{k}} = \frac{2\pi}{\hbar} \vert \langle u_s(\vec{k}) \vert \hat{V} \vert \psi_s(\vec{k}')\rangle \vert^{2} \delta \big( \epsilon_{s}(\vec k) - \epsilon_{s}(\vec k') \big), 
\end{equation} 
where $|\psi_s(\vec{k}')\rangle$ is the eigenstate of the full Hamiltonian $\hat{H}^{(0)}_s + \hat{V}$, and $|u_s(\vec{k})\rangle$ is the eigenstate of the bare Hamiltonian without impurity, $\hat{H}^{(0)}_s$. Applying the Lippmann-Schwinger self-consistency relation, $|\psi_s(\vec{k}')\rangle$ reads as $|\psi_s(\vec{k}')\rangle = |u_s(\vec{k}')\rangle + \frac{1}{\epsilon_s(\vec{k}') - \hat{H}^{(0)}_s+i\eta}\hat{V}|\psi_s(\vec{k}')\rangle$. Expanding iteratively produces 
\begin{equation}
\begin{aligned}
\label{eq:inner-product}
\langle u_s(\vec{k}) \vert \hat{V} \vert \psi_s(\vec{k}')\rangle = \langle u_s(\vec{k}) \vert \hat{V} \vert u_s(\vec{k}')\rangle + \sum_{\vec{k}''} \frac{\langle u_s(\vec{k}) \vert \hat{V} \vert u_s(\vec{k}'')\rangle \langle u_s(\vec{k}'') \vert \hat{V} \vert u_s(\vec{k}')\rangle }{\epsilon_s(\vec{k}') - \epsilon_s(\vec{k}'')+i\eta} + \cdot\,\cdot\,\cdot \\ + \sum_{\vec{k}''}\cdot\cdot\cdot\sum_{\vec{k}^{('n)}}\frac{\langle u_s(\vec{k})\vert \hat{V} |u_s(\vec{k}'')\rangle\langle u_s(\vec{k}'')|\cdot\cdot\cdot|u_s(\vec{k}^{('n)})\rangle\langle u_s(\vec{k}^{('n)}) \vert \hat{V} | u_s(\vec{k}')\rangle}{[\epsilon_s(\vec{k}') - \epsilon_s(\vec{k}'')+i\eta] \cdot\cdot\cdot [\epsilon_s(\vec{k}') - \epsilon_s(\vec{k}^{('n)})+i\eta] } +\cdots,
\end{aligned}
\end{equation}
where we have introduced $\vec{k}^{('n)}$ so that $\vec{k}^{('1)} = \vec{k}''$ for $n=1$, $\vec{k}^{('2)} = \vec{k}'''$ for $n=2$, and so on. Importantly, the energy eigenstates in a {\it PT}-symmetric material obey Eq.~(\ref{eq:PTsymmetry}) of the main text; this constrains the spin-dependent rates in Eq.~(\ref{eq:inner-product}). To see this explicitly, consider the matrix elements of the scalar disorder potential $\hat{V}(\bm{r}) = V_{0} \sum_{j} \delta \left( \vec{r} - \vec{r}_{j} \right)$. Applying Eq.~(\ref{eq:PTsymmetry}) of the main text yields $\langle u_s(\vec{k}) \vert \hat{V} \vert u_s(\vec{k}')\rangle = \langle u_{\bar{s}}(\vec{k}') \vert \hat{V} \vert u_{\bar{s}}(\vec{k})\rangle$. This also holds for generic $PT$-symmetric $\hat{V}$'s. As a result, we find 
\begin{equation}
\begin{aligned}
\label{eq:inner-product-2}
\langle u_s(\vec{k}) \vert \hat{V} \vert \psi_s(\vec{k}')\rangle = \langle u_{\bar{s}}(\vec{k}') \vert \hat{V} \vert u_{\bar{s}}(\vec{k})\rangle + \sum_{\vec{k}''} \frac{\langle u_{\bar{s}}(\vec{k}'') \vert \hat{V} \vert u_{\bar{s}}(\vec{k})\rangle \langle u_{\bar{s}}(\vec{k}') \vert \hat{V} \vert u_{\bar{s}}(\vec{k}'')\rangle }{\epsilon_{\bar{s}}(\vec{k}') - \epsilon_{\bar{s}}(\vec{k}'')+i\eta} + \cdot\,\cdot\,\cdot \\ + \sum_{\vec{k}''}\cdot\cdot\cdot\sum_{\vec{k}^{('n)}}\frac{\langle u_{\bar{s}}(\vec{k}'')\vert \hat{V} |u_{\bar{s}}(\vec{k})\rangle \cdot\cdot\cdot \langle u_{\bar{s}}(\vec{k}') \vert \hat{V} | u_{\bar{s}}(\vec{k}^{('n)})\rangle}{[\epsilon_{\bar{s}}(\vec{k}') - \epsilon_{\bar{s}}(\vec{k}'')+i\eta] \cdot\cdot\cdot [\epsilon_{\bar{s}}(\vec{k}') - \epsilon_{\bar{s}}(\vec{k}^{('n)})+i\eta] } + \cdots = \langle u_{\bar{s}}(\vec{k}') \vert \hat{V} \vert \psi_{\bar{s}}(\vec{k})\rangle ,
\end{aligned}
\end{equation}
where in obtaining the last equality we have recalled that the scattering process we are concerned with is elastic so that $\epsilon_{\bar{s}}(\vec{k}') = \epsilon_{\bar{s}}(\vec{k})$. Substituting Eq.~(\ref{eq:inner-product-2}) into Eq.~(\ref{eq:gold}) immediately produces
\begin{equation}
W^\uparrow_{\vec{k}\rightarrow\vec{k}'} = W^\downarrow_{\vec{k}'\rightarrow\vec{k}}
\end{equation}
reproducing Eq.~(\ref{eq:rate}) of the main text. Importantly, by examining the symmetric ($w^S_{s,\vec{k},\vec{k}'} = [W^s_{\vec{k'}\rightarrow\vec{k}} + W^s_{\vec{k}\rightarrow\vec{k}'}]/2$) and antisymmetric (i.e. skew) ($w^A_{s,\vec{k},\vec{k}'} = [W^s_{\vec{k}'\rightarrow\vec{k}} - W^s_{\vec{k}\rightarrow\vec{k}'}]/2$) parts of the scattering rate above, we obtain 
\begin{equation}
\label{sym-skew-rates}
w^S_{\uparrow,\vec{k},\vec{k}'} = w^S_{\downarrow,\vec{k},\vec{k}'} \quad {\rm and}\quad w^A_{\uparrow,\vec{k},\vec{k}'} = -w^A_{\downarrow,\vec{k},\vec{k}'}. 
\end{equation} 
Note that the total scattering rate for either $\uparrow$ or $\downarrow$ spins is always positive definite; in contrast, the {\it anti-symmetric} part of the rate can have opposite signs for $\uparrow$ and $\downarrow$ processes. For a discussion of the effect of spin-non-conserving scattering, see the end of the section ``nonlinear responses" below.

\subsection{Leading order symmetric and antisymmetric scattering rates}
While the above analysis is valid at all orders in $V$, as a concrete example, we now discuss the leading-order symmetric and asymmetric scattering rates within the description above. Plugging Eq.~(\ref{eq:inner-product}) into Eq.~(\ref{eq:gold}), taking $\eta\rightarrow 0^{+}$, we write the scattering rate in Eq.~(\ref{eq:gold}) to third order in $V$ as~\cite{Luttinger1958,Sinitsyn2006,Nagaosa2010} 
\begin{equation}
\begin{aligned}
\label{eq:scatteringOV3}
W^{s}_{\vec{k}'\rightarrow\vec{k}} = & \frac{2\pi}{\hbar} \bigg[ \vert \langle u_{{s}}(\vec{k}) \vert \hat{V} \vert u_{{s}}(\vec{k}')\rangle \vert^{2} + 2 \sum_{\vec{k}''}\mathrm{Re}\frac{ \langle u_{{s}}(\vec{k}) \vert \hat{V} \vert u_{{s}}(\vec{k}'')\rangle \langle u_{{s}}(\vec{k}'') \vert \hat{V} \vert u_{{s}}(\vec{k}')\rangle \langle u_{{s}}(\vec{k}') \vert \hat{V} \vert u_{{s}}(\vec{k})\rangle }{\epsilon_s(\vec{k}') - \epsilon_s(\vec{k}'') } + \\
& 2 \pi \sum_{\vec{k}''}  \mathrm{Im} \left( \langle u_{{s}}(\vec{k}) \vert \hat{V} \vert u_{{s}}(\vec{k}'')\rangle \langle u_{{s}}(\vec{k}'') \vert \hat{V} \vert u_{{s}}(\vec{k}')\rangle \langle u_{{s}}(\vec{k}') \vert \hat{V} \vert u_{{s}}(\vec{k})\rangle \right) \delta \big( \epsilon_{s}(\vec k') - \epsilon_{s}(\vec k'') \big) \bigg] \delta \big( \epsilon_{s}(\vec k) - \epsilon_{s}(\vec k') \big) + \mathcal{O}(V^4)  ,
\end{aligned}
\end{equation}
where we have applied $\langle u_{{s}}(\vec{k}) \vert \hat{V} \vert u_{{s}}(\vec{k}')\rangle =  \left( \langle u_{{s}}(\vec{k}') \vert \hat{V} \vert u_{{s}}(\vec{k})\rangle \right)^{*}$. Note that averaging over impurity distributions ($\vec{r}_{j}$) is implied. Notably, the symmetric/antisymmetric properties of each of the terms of Eq.~(\ref{eq:scatteringOV3}) can be discerned directly by switching $\vec k$ and $\vec k'$: the first and the second terms in the square bracket are symmetric under $\vec k \longleftrightarrow \vec k'$, while the third is antisymmetric under $\vec k \longleftrightarrow \vec k'$. 

The first term in the square bracket of Eq.~(\ref{eq:scatteringOV3}) contains the leading order contribution to symmetric scattering rate,
\begin{equation}
w^S_{s,\vec{k},\vec{k}'}  =  \frac{2\pi}{\hbar} n_{i} V_{0}^{2}  \vert \langle u_{{s}}(\vec{k}) \vert  u_{{s}}(\vec{k}')\rangle \vert^{2} \delta \big( \epsilon_{s}(\vec k) - \epsilon_{s}(\vec k') \big) ,
\end{equation}
where $n_{i}$ is the density of scalar impurities. 

Similarly, we obtain the third-order (antisymmetric/skew) scattering from Eq.~(\ref{eq:scatteringOV3}) \cite{Luttinger1958,Sinitsyn2006,Nagaosa2010},
\begin{equation}
\begin{aligned}
w^A_{s,\vec{k},\vec{k}'} = \frac{4\pi^{2}}{\hbar} n_{i} V_{0}^{3} \sum_{\vec{k}''} \mathrm{Im} \big[ L_{s}(\vec{k},\vec{k}'',\vec{k}') \big] \delta \big( \epsilon_{s}(\vec k') - \epsilon_{s}(\vec k'') \big) \delta \big( \epsilon_{s}(\vec k) - \epsilon_{s}(\vec k') \big),
\end{aligned}
\end{equation}
where $L_s(\vec{k},\vec{k}'', \vec{k}') = \langle u_s(\vec{k}) | u_s (\vec{k}'') \rangle \langle u_s(\vec{k}'')| u_s(\vec{k}')\rangle\langle u_s(\vec{k}')|u_s(\vec{k})\rangle$ is the gauge invariant Wilson loop. For {\it PT} partners, $L_{s}(\vec{k},\vec{k}'',\vec{k}') = \left( L_{\bar{s}}(\vec{k},\vec{k}'',\vec{k}') \right)^{*}$, which means that skew-scattering contribution to the total scattering rate of opposite spin polarization here has opposite signs, in line with the conclusions of the previous section. 

\subsection{Nonlinear response and the kinetic equation}

In this section, we review the standard treatment of the kinetic equation for a spatially uniform system including both symmetric as well as antisymmetric (skew) scattering rates. We then connect it with second-order nonlinear response. In what follows, we will concentrate on the intraband/semiclassical limit; see the end of the section for a short discussion of interband effects.

\subsubsection{Distribution function and kinetic equation}

The kinetic equation reads as 
\begin{equation}
\label{A:eq:Boltzmann}
\frac{\partial  }{\partial t} f_s(\vec k,t) - \frac{e}{\hbar} \boldsymbol{\mathcal{E}} (t)  \cdot \partial_{\vec{k} }f_s(\vec k,t) = \mathcal{I}\{ f_s(\vec k,t) \},
\end{equation} 
where $\boldsymbol{\mathcal{E}}(t) = \vec E e^{i\omega t} + c.c.$, $s = \{ \uparrow, \downarrow\}$, and the collision integral \cite{Kohn-Luttinger1957,Sinitsyn2007,Nagaosa2010,XCXie2019,Konig2019,LiangFu2020}
\begin{equation}
\label{eq:collision-int-o}
\mathcal{I}\{ f_s(\vec k,t)  \} = - \sum_{\vec{k}^{\prime}} \left\{ W_{\vec k \to \vec k'}^{s} f_s(\vec k,t)  - W_{\vec k' \to \vec k}^{s} f_s( \vec{k}^{\prime} ,t) \right\} .
\end{equation}
Eq.(\ref{A:eq:Boltzmann}) is solved in the standard perturbative fashion by writing~\cite{XCXie2019,LiangFu2020},
\begin{equation}
f_{s} (\vec k,t) = f_{0}(\vec k) + \sum _{\ell, m} f_{\ell ,s}^{(m)} (\vec k,t), 
\end{equation}
where $f_{0}(\vec k)$ is the equilibrium distribution function, which is the same for both $\uparrow$ and $\downarrow$ spins due to {\it PT} symmetry, subscript $\ell$ indicates order in $\vec E$, and the superscript $m$ denotes the power dependence on skew-scattering rate; $m=0$ corresponds to purely symmetric scattering. Accordingly, we can decompose the kinetic equation in orders of $\ell$ and $m$, separating the equation into one part involving purely symmetric scattering ($m=0$),
\begin{equation}
\begin{aligned}
\label{eq:A-BE-f1(0)}
\frac{\partial }{\partial t} f_{1 ,s}^{(0)} (\vec k,t) - \frac{e}{\hbar} \boldsymbol{\mathcal{E}} (t) \cdot \partial_{\vec{k} }f_{0}(\vec k)  = & - \sum_{\vec{k}^{\prime} } w^{({\rm S})}_{s, \vec k, \vec k'} \left( f_{1 ,s}^{(0)} (\vec k,t) - f_{1 ,s}^{(0)} (\vec{k}^{\prime},t) \right) ,
\end{aligned}
\end{equation} 
\begin{equation}
\begin{aligned}
\label{eq:A-BE-f2(0)}
\frac{\partial }{\partial t} f_{2 ,s}^{(0)} (\vec k,t) - \frac{e}{\hbar} \boldsymbol{\mathcal{E}} (t) \cdot \partial_{\vec{k} }f_{1,s}^{(0)} (\vec k, t)  = & - \sum_{\vec{k}^{\prime} } w^{({\rm S})}_{s, \vec k, \vec k'} \left( f_{2 ,s}^{(0)} (\vec k,t) - f_{2 ,s}^{(0)} (\vec{k}^{\prime},t) \right) ,
\end{aligned}
\end{equation} 
and another involving antisymmetric (skew) scattering ($m=1$),
\begin{equation}
\begin{aligned}
\label{eq:A-BE-f1(1)}
\frac{\partial  }{\partial t} f_{1 ,s}^{(1)} (\vec k,t) = & - \sum_{\vec{k}^{\prime}} \bigg\{ w^{\mathrm{S}}_{s, \vec{k}^{\prime}, \vec{k} } \left( f_{1 ,s}^{(1)} (\vec k,t)  - f_{1 ,s}^{(1)} (\vec{k}^{\prime},t)  \right)  - w^{\mathrm{A}}_{s, \vec{k}, \vec{k}^{\prime} } f_{1 ,s}^{(0)} (\vec{k}^{\prime},t)  \bigg\} ,
\end{aligned}
\end{equation} 
\begin{equation}
\begin{aligned}
\label{eq:A-BE-f2(1)}
\frac{\partial }{\partial t} f_{2 ,s}^{(1)} (\vec k,t) - \frac{e}{\hbar} \boldsymbol{\mathcal{E}} (t) \cdot \partial_{\vec{k} } f_{1 ,s}^{(1)} (\vec k,t) = & - \sum_{\vec{k}^{\prime} } \bigg\{ w^{\mathrm{S}}_{s, \vec{k}^{\prime}, \vec{k} } \left( f_{2 ,s}^{(1)} (\vec k,t)  - f_{2 ,s}^{(1)} (\vec{k}^{\prime},t) \right)  - w^{\mathrm{A}}_{s, \vec{k}, \vec{k}^{\prime} } f_{2 ,s}^{(0)} (\vec{k}^{\prime},t)  \bigg\} .
\end{aligned}
\end{equation} 
where in obtaining Eq.~(\ref{eq:A-BE-f1(1)}) and (\ref{eq:A-BE-f2(1)}) we have noted $W_{\vec k' \to \vec k}^{s} =  w^{\mathrm{(S)}}_{s, \vec{k} , \vec{k}^{\prime}}  +  w^{\mathrm{(A)}}_{s, \vec{k}, \vec{k}^{\prime}}$ and $W_{\vec k \to \vec k'}^{s} =  w^{\mathrm{(S)}}_{s, \vec{k} , \vec{k}^{\prime}}  -  w^{\mathrm{(A)}}_{s, \vec{k}, \vec{k}^{\prime}}$, and we have simplified the expression with probability conservation, $\sum_{\vec{k}^{\prime} } w^{\mathrm{A}}_{s, \vec{k}, \vec{k}^{\prime} }  = 0$~\cite{Konig2019,LiangFu2020}. In what follows we concentrate on the skew-scattering induced deviation to the distribution function to lowest order, $m=1$.

To proceed with the analysis, we adopt the standard relaxation time approximation for symmetric scattering~\cite{XCXie2019,Konig2019,ashcroft1976}, 
\begin{equation}
\label{eq:A-RTA}
\sum_{\vec{k}^{\prime}} w^{\mathrm{S}}_{s, \vec{k}^{\prime}, \vec{k} } \left( f_{\ell ,s}^{(m)} (\vec k,t) - f_{\ell ,s}^{(m)} (\vec{k}^{\prime},t) \right) \approx \frac{1}{\tau^{s} } f_{\ell ,s}^{(m)} (\vec k,t) , \quad \left( \tau^s \right)^{-1} = \langle \sum_{\vec k'} w^{S}_{s, \vec k', \vec k} \left( 1 - \cos\theta_{\vec{v} \vec{v}' } \right) \rangle, 
\end{equation}
where $\theta_{\vec{v} \vec{v}' }$ is the angle between $\vec{v}(\vec{k})$ and $\vec{v}(\vec{k}')$, and $\langle \rangle$ indicates averaging over the energy contour. This approximation has been widely used to successfully describe nonlinear responses from a semiclassical kinetic equation-like treatment, see e.g., extensive discussion in Ref.~\cite{Konig2019,ashcroft1976}.

Symmetric scattering in Eqs.(\ref{eq:A-BE-f1(0)}) and (\ref{eq:A-BE-f2(0)}) leads to conventional (symmetric scattering induced) deviation to the distribution function,
\begin{equation}
\label{eq:A-BE-f-0-1}
f_{1,s}^{(0)} (\vec k,t) = \frac{e }{\hbar}  \tau_\omega^s e^{i\omega t} \vec E\cdot \partial_{\vec{k} }f_{0}(\vec k) +c.c., 
\end{equation}
\begin{equation}
\label{eq:A-BE-f-0-2}
f_{2,s}^{(0)} (\vec k,t) = \frac{e^{2} }{\hbar^{2}} \left( \tau_{2\omega}^s e^{2 i\omega t} \vec E \cdot \partial_{\vec{k} } + \tau^{s} \vec E^{*} \cdot \partial_{\vec{k} } \right) \Big[ \tau_\omega^s \vec E\cdot \partial_{\vec{k} }f_{0}(\vec k) \Big] +c.c.,
\end{equation}
while asymmetric scattering in Eqs.(\ref{eq:A-BE-f1(1)}) and (\ref{eq:A-BE-f2(1)}) leads to skew-scattering part of the distribution functions, 
\begin{equation}
\label{eq:A-BE-f-1-1}
f_{1,s}^{(1)} (\vec k,t) = \frac{e }{\hbar}  \left( \tau_\omega^s \right)^{2} e^{i\omega t} \sum_{\bm{k}^{\prime}} w^{\mathrm{A}}_{s, \vec{k}, \vec{k}^{\prime} } \vec E\cdot \partial_{\vec{k}^{\prime} }f_{0}(\vec k^{\prime}) +c.c.,
\end{equation}
\begin{equation}
\begin{aligned} 
\label{eq:A-BE-f-1-2}
f_{2,s}^{(1)} (\vec k,t) = & \frac{e^{2} }{\hbar^{2}} \left( \tau_\omega^s \right)^{2} \left( \tau_{2 \omega}^s e^{2 i\omega t} \vec{E} \cdot \partial_{\vec{k} } + \tau^s \vec{E}^{*} \cdot \partial_{\vec{k} } \right) \sum_{\vec{k}^{\prime}} w^{\mathrm{A}}_{s, \vec{k}, \vec{k}^{\prime} } \vec{E} \cdot \partial_{\vec{k}^{\prime} } f_{0}(\vec k^{\prime}) \\
& + \frac{e^{2} }{\hbar^{2}} \tau_\omega^s \sum_{\vec{k}^{\prime}} w^{\mathrm{A}}_{s, \vec{k}, \vec{k}^{\prime} } \big[ \left(\tau_{2 \omega}^s \right)^{2} e^{2 i\omega t} \vec{E} \cdot \partial_{\vec{k}^{\prime} } + \left(\tau^s \right)^{2} \vec{E}^{*} \cdot \partial_{\vec{k}^{\prime} } \big] \vec{E} \cdot \partial_{\vec{k}^{\prime} } f_{0}(\vec k^{\prime}) + c.c., 
\end{aligned} 
\end{equation}
where $\tau_\omega^s = \tau^s/(1+i\omega \tau^s)$; similarly, $\tau_{2\omega}^s = \tau^s/(1+i2\omega \tau^s)$. Note that in for a $PT$-symmetric system, $\tau^\uparrow = \tau^\downarrow = \tau$. As a result, the term $\tau_\omega^s$ can be simply written as $\tilde\tau_\omega = \tau/(1+i\omega \tau)$ as used in the main text. These distribution function are used in the main text (as described below) to obtain the nonlinear susceptibilities. Throughout the supplementary information, and in Table \ref{Table:SM-T-Classification} and~\ref{Table:SM-PT-Classification}, we keep full spin dependence $s$ so as to enable to describe situations beyond $PT$-symmetric systems.

\subsubsection{Nonlinear responses}
Semiclassically, the current can be expressed as the combination of the velocity of the charge carriers and the distribution function (obtained above) \cite{QianNiu2014,DiXiao2021,ShengyuanYang2021}, 
\begin{equation}
\label{SI:eq:j}
\vec{j} (t) = - e \sum_{\vec{k}, s} \left( \vec{v} (\vec k) + \frac{e}{\hbar} \boldsymbol{\mathcal{E}} (t) \times \bar{\boldsymbol{\Omega}}_{s} (\vec k) \right) f_{s} (\vec k,t),
\end{equation}
where the modified Berry curvature $\bar{\boldsymbol{\Omega}}_s (\vec k)$ contains both intrinsic Bloch band Berry curvature $\boldsymbol{\Omega_{s}} (\vec k)$ and the field-induced Berry curvature \cite{QianNiu2014,DiXiao2021,ShengyuanYang2021},
\be
\bar{\boldsymbol{\Omega}}_{s} (\vec k) = \boldsymbol{\Omega}_{s} (\vec k) + \nabla_{\vec k} \times \boldsymbol{\mathcal{G}} (\vec k)\boldsymbol{\mathcal{E}} (t). 
\ee
The Berry connection polarizability tensor $\boldsymbol{\mathcal{G}} (\vec k)$ contains interband Berry connection. For band $n$ \cite{QianNiu2014,DiXiao2021,ShengyuanYang2021},
\begin{equation}
{\mathcal{G}}_{ab} (\vec k)  = 2 e {\rm Re} \sum_{n'\neq n} \frac{ A_a^{nn'} (\vec k) A_b^{n'n} (\vec k)}{ \epsilon_{n}(\vec k) - \epsilon_{n'}(\vec k)} ,
\end{equation}
where $n'$ is the index of another band, and band numbered $n'$ and $n$ are not degenerate. The interband Berry connection is defined as $ A_a^{nn'} = \la u_{n} (\vec k) | i\partial_{k_a} u_{n^{\prime}}(\vec k)  \ra$. While the intrinsic Berry curvature $\boldsymbol{\Omega_{s}} (\vec k)$ is {\it PT}-odd, ${\mathcal{G}}$ is {\it PT}-even \cite{DiXiao2021,ShengyuanYang2021}. 

Generically speaking, Eq.~(\ref{SI:eq:j}) contains four kinds of second-order nonlinear Hall response here: the intrinsic, the Berry curvature dipole, the conventional skew-scattering, and the anomalous skew-scattering nonlinear Hall effects. We note that combination of the velocity $\vec{v} (\vec k)$ and the symmetric scattering induced nonlinear distribution function $f_{2,s}^{(0)}$ leads to another nonlinear response: nonlinear ``Drude'' response \cite{Yanase2020,DiXiao2021}. However, the indices of its susceptibility is totally symmetric, and being purely ohmic in nature \cite{Souza2022}, does not contribute to a nonlinear Hall effect. In what follows, we focus only on effects that can lead to nonlinear Hall type responses. 

These four second-order responses can be categorized into two groups, and can be expressed succinctly as
\begin{equation}
j_{a} (t) = \mathrm{Re} \left( j_{a}^{0} +  j_{a}^{2\omega} e^{2 i \omega t} \right),
\end{equation}
\begin{equation}
j_{a}^{0} = \left( \chi_{abc} + \chi_{abc}^{\mathrm{T}, 0} \right) E_{b}^{*} E_{c} , \quad j_{a}^{2\omega} = \left( \chi_{abc} + \chi_{abc}^{\mathrm{T}, 2\omega} \right) E_{b} E_{c},
\end{equation}
where $\chi_{abc}$ corresponds to the group which is {\it PT}-even and {\it T}-odd, including the anomalous skew-scattering nonlinear Hall (ASN)  effect and the intrinsic anomalous nonlinear Hall (INH) effect,
\begin{equation}
\chi_{abc} = \chi_{abc}^{\mathrm{ASN}} + \chi_{abc}^{\mathrm{INH}}.
\end{equation}
The anomalous skew-scattering comes from the combination of the anomalous velocity induced by intrinsic Berry curvature $\boldsymbol{\Omega_{s}} (\vec k)$ and the distribution function  $f_{1 ,s}^{(1)} (\vec k,t)$. The ASN susceptibility is 
\begin{equation}
\label{eq:SI-chi-ASN}
\chi_{abc}^{{\rm ASN}} = \frac{2 e^3\varepsilon_{adb}}{\hbar^2}\sum_{\vec k, \vec k', s} \Omega^{s}_{d}(\vec k) \tau_{s,\omega}^{2} w^A_{s,\vec k,\vec k'} \left[\frac{\partial f_0 (\vec k')}{\partial \vec k'}\right]_c,
\end{equation}
where $\tau_{s,\omega}^{2} = \left( \tau^{s}_{\omega} \right)^{2}$. Flipping spin polarization $s$ would change the sign of both the Berry curvature $\Omega$ and the skew-scattering rate $w^A_{s,\vec k,\vec k'}$. Thus, this response is {\it PT}-even. Note that for a $PT$-symmetric system, $\tau^\uparrow = \tau^\downarrow = \tau$. As a result, the term $\tau_\omega^s$ can be simply written as $\tilde\tau_\omega = \tau/(1+i\omega \tau)$ reproducing the ASN susceptibility in the main text in Eq.~(\ref{eq:ASN}). We retain full $s$ dependence in the supplementary information for clarity.

The intrinsic anomalous nonlinear Hall effect is the result of field-induced Berry curvature and equilibrium distribution function $f_{0} (\vec k)$ \cite{QianNiu2014,DiXiao2021,ShengyuanYang2021},
\begin{equation}
\chi_{abc}^{\mathrm{INH}} = 4 e^{3} \sum_{\vec k, s} \sum_{n'\neq n} {\rm Re} \frac{ v^{n}_{a} A_{b}^{nn'} A_{c}^{n'n} - v^{n}_{b} A_{a}^{nn'} A_{c}^{n'n} }{ \epsilon_{n} - \epsilon_{n'} } \frac{\partial f_{0} }{\partial \epsilon_{n } }.
\label{eq:INHchiSI}
\end{equation}
The INH susceptbility could also be expressed with the Berry connection polarizability $\mathcal{G}$, $\chi_{abc}^{\mathrm{INH}} = 2 e^{2} \sum_{\vec k, s} \left( v_{a} \mathcal{G}_{bc} - v_{b} \mathcal{G}_{ac} \right) \frac{\partial f_{0} }{\partial \epsilon_{n } }$. As stated earlier, $\mathcal{G}$ is {\it PT}-even. So is the INH term. In writing the expression in Eq.~(\ref{eq:INHchiSI}), we have focussed on the intraband/semiclssical limit. A discussion of interband effects can be found at the end of this section.

The other group is {\it PT}-odd and {\it T}-even, and is contained in $\chi_{abc}^{\mathrm{T}, 0}$ and $\chi_{abc}^{\mathrm{T}, 2\omega}$, $\chi_{abc}^{\mathrm{T}, 0} = \chi_{abc}^{\mathrm{BCD}} + \chi_{abc}^{\mathrm{SS}, 0}$, $\chi_{abc}^{\mathrm{T}, 2\omega} = \chi_{abc}^{\mathrm{BCD}} + \chi_{abc}^{\mathrm{SS}, 2\omega}$. The Berry curvature dipole (BCD) part is made of the intrinsic Berry curvature and distribution function induced by symmetric scattering $f_{1,s}^{(0)} (\vec k,t)$ \cite{Sodemann2015},
\begin{equation}
\chi_{abc}^{\mathrm{BCD}} = 2 \frac{e^{3} }{ \hbar^{2} } \varepsilon_{abd} \sum_{\vec k, s} \tau^{s}_{\omega} f_{0} \left[\frac{\partial \Omega_{d}^{s} (\vec k)}{\partial \vec k}\right]_c .
\end{equation}
As the sign of Berry curvature flips when spin $s$ is flipped, the BCD contribution is {\it PT}-odd. The conventional skew-scattering (SS) contribution comes from the combination of $\vec{v} (\vec k)$ and  $f_{2,s}^{(1)}$,
\begin{equation}
\chi_{abc}^{\mathrm{SS}, 0} = - 2 \frac{e^{3} }{ \hbar^{2} } \sum_{\vec k, s} {v}_{a} \bigg\{ \tau_{s,\omega}^{2} \tau^{s} \frac{\partial }{\partial {k}_{b} } \sum_{ \vec{k}^{\prime} } w^{A}_{s,\vec k,\vec k'} \left[\frac{\partial f_0 (\vec k')}{\partial \vec k'}\right]_c + \tau^{s}_{\omega} \left( \tau^{s} \right)^{2} \sum_{ \vec{k}^{\prime} } w^{A}_{s,\vec k,\vec k'}  \left[\frac{\partial }{\partial \vec k'} \left(\frac{\partial f_0 (\vec k')}{\partial \vec k'}\right)_c \right]_b \bigg\} ,
\end{equation}
\begin{equation}
\chi_{abc}^{\mathrm{SS}, 2\omega} = - 2 \frac{e^{3} }{ \hbar^{2} } \sum_{\vec k, s} {v}_{a}  \bigg\{ \tau_{s,\omega}^{2} \tau^{s}_{2\omega} \frac{\partial }{\partial {k}_{b} } \sum_{ \vec{k}^{\prime} } w^{A}_{s,\vec k,\vec k'} \left[\frac{\partial f_0 (\vec k')}{\partial \vec k'}\right]_c +  \tau^{s}_{\omega} \left( \tau^{s}_{2\omega} \right)^{2} \sum_{ \vec{k}^{\prime} } w^{A}_{s,\vec k,\vec k'}\left[\frac{\partial }{\partial \vec k'} \left(\frac{\partial f_0 (\vec k')}{\partial \vec k'}\right)_c \right]_b \bigg\} .
\end{equation}
In the same fashion, since the skew-scattering rate has opposite signs for the two spin polarizations $\uparrow$ and $\downarrow$, the conventional skew-scattering nonlinearity is also {\it PT}-odd. 

Interestingly, the form of the ASN response remains intact even if one considers weak spin flipping due to impurity scattering, where spin flipping can be described by an effective spin-flipping relaxation time $\tau_{\mathrm{SF}}$. When such spin-flipping is included, the factor $\tilde\tau_{\omega}^{2}$ in Eq.~(\ref{eq:ASN}) in the main text [equivalently $\tilde\tau_{s,\omega}^{2}$ in Eq.~(\ref{eq:SI-chi-ASN}) of the supplementary information] is renormalized and is replaced by $\tilde\tau_{\omega} \tilde{\tau}_{\omega,{\rm renorm}}$, where $\tilde{\tau}_{\omega, \rm renorm} = {\tau}_{\rm renorm}/(1+i\omega {\tau}_{\rm renorm})$ and ${\tau}_{{\rm renorm}} = \tau\tau_{\mathrm{SF}}/(2\tau+\tau_{\mathrm{SF}})$. Crucially, ASN persists even with spin flipping, with the form of the susceptibility fully intact.

Lastly, we note that here we have concentrated on the semiclassical limit where $\omega  \ll \epsilon_n - \epsilon_m$ (the characteristic scale of interband transitions). However, for larger $\omega$ when interband transitions become important, a range of other nonlinearities can become activated~\cite{Ahn-Nagaosa2020,Watanabe2021,Bhalla2023}. For instance, in {\it PT}-symmetric systems, these can include the intrinsic Fermi surface effect~\cite{Watanabe2021}, as well as nonlinearities that result from interband transition absorption effects such as a linear polarized light induced injection photocurrent (sometimes referred to as the linear injection current~\cite{Ahn-Nagaosa2020,Watanabe2021}) and a circularly polarized light induced chiral photocurrent~\cite{Ahn-Nagaosa2020,Watanabe2021}. The last of these is sometimes referred to as the gyration current~\cite{Watanabe2021} (it is also known as the circular shift photocurrent elsewhere in the literature, e.g.,~\cite{Ahn-Nagaosa2020}). These nonlinearities have a characteristic frequency dependence that tracks energies corresponding to interband transitions~\cite{Ahn-Nagaosa2020,Watanabe2021} enabling to clearly distinguish from the intraband phenomena discussed above.

\begin{table}
\begin{center}
\begin{tabular}{l c c c}
\hline\hline
Nonlinear Hall effects & $\chi_{abc}$ & $T$   \\
\hline 
Berry curvature dipole (BCD) & $2 \frac{e^{3} }{ \hbar^{2} } \varepsilon_{abd} \sum_{\vec k, s} {\color{ForestGreen} \tau^{s}_{\omega} f_{0}} \left[\frac{\partial {\color{red} \Omega_{d}^{s} (\vec k)}}{\partial{\color{red}  \vec k}}\right]_c$ & ${\color{ForestGreen}+}$  \\
Intrinsic (INH)                      & $4 e^{3} \sum_{\vec k, s} \sum_{n'\neq n} {\rm Re} \frac{ {\color{red} v^{n}_{a}} {\color{ForestGreen} A_{b}^{nn'} A_{c}^{n'n}} - {\color{red} v^{n}_{b}} {\color{ForestGreen}A_{a}^{nn'} A_{c}^{n'n}} }{ {\color{ForestGreen}\epsilon_{n}} - {\color{ForestGreen}\epsilon_{n'}} } \frac{\partial {\color{ForestGreen}f_{0}} }{\partial {\color{ForestGreen}\epsilon_{n }} }$  & ${\color{red}-}$  \\
Conventional skew-scattering    &  $- 2 \frac{e^{3} }{ \hbar^{2} } \sum_{\vec k, s} {\color{red}{v}_{a}} \bigg\{ {\color{ForestGreen}\tau_{s,\omega}^{2} \tau^{s}} \frac{\partial }{\partial {\color{red}{k}_{b}} } \sum_{ \vec{k}^{\prime} } {\color{red}w^{A}_{s,\vec k,\vec k'}} \left[\frac{\partial {\color{ForestGreen}f_0 (\vec k')}}{\partial {\color{red}\vec k'}}\right]_c + {\color{ForestGreen}\tau^{s}_{\omega} \left( \tau^{s} \right)^{2}} \sum_{ \vec{k}^{\prime} } {\color{red}w^{A}_{s,\vec k,\vec k'}}  \left[\frac{\partial }{\partial {\color{red}\vec k'}} \left(\frac{\partial {\color{ForestGreen}f_0 (\vec k')}}{\partial {\color{red}\vec k'}}\right)_c \right]_b \bigg\}$     & ${\color{ForestGreen}+}$  \\
Anomalous skew-scattering (ASN)   & $\frac{2 e^3\varepsilon_{adb}}{\hbar^2}\sum_{\vec k, \vec k', s} {\color{red}\Omega^{s}_{d}(\vec k)} {\color{ForestGreen}\tau_{s,\omega}^{2}} {\color{red}w^A_{s,\vec k,\vec k'}} \left[\frac{\partial {\color{ForestGreen}f_0 (\vec k')}}{\partial {\color{red}\vec k'}}\right]_c$ & ${\color{red}-}$  \\  
\hline \hline
\end{tabular}
\end{center}
\caption{Intraband nonlinear Hall responses and time-reversal symmetry. Green color indicates that the quantity is even under $T$ symmetry, while red color means it is $T$-odd. Similarly, ${\color{ForestGreen}+}$ indicates the response is even (i.e. allowed by symmetry), $-$ means it is odd (i.e. forbidden by symmetry).}
\label{Table:SM-T-Classification}
\end{table}

\begin{table}
\begin{center}
\begin{tabular}{l c c c}
\hline\hline
Nonlinear Hall effects & $\chi_{abc}$ & $PT$  \\
\hline 
Berry curvature dipole (BCD) & $2 \frac{e^{3} }{ \hbar^{2} } \varepsilon_{abd} \sum_{\vec k, s} {\color{ForestGreen} \tau^{s}_{\omega} f_{0}} \left[\frac{\partial {\color{red} \Omega_{d}^{s} (\vec k)}}{\partial{\color{ForestGreen}  \vec k}}\right]_c$ & ${\color{red}-}$          \\
Intrinsic (INH)                      & $4 e^{3} \sum_{\vec k, s} \sum_{n'\neq n} {\rm Re} \frac{ {\color{ForestGreen} v^{n}_{a}} {\color{red} A_{b}^{nn'} A_{c}^{n'n}} - {\color{ForestGreen} v^{n}_{b}} {\color{red}A_{a}^{nn'} A_{c}^{n'n}} }{ {\color{ForestGreen}\epsilon_{n}} - {\color{ForestGreen}\epsilon_{n'}} } \frac{\partial {\color{ForestGreen}f_{0}} }{\partial {\color{ForestGreen}\epsilon_{n }} }$  & ${\color{ForestGreen}+}$  \\
Conventional skew-scattering    &  $- 2 \frac{e^{3} }{ \hbar^{2} } \sum_{\vec k, s} {\color{ForestGreen}{v}_{a}} \bigg\{ {\color{ForestGreen}\tau_{s,\omega}^{2} \tau^{s}} \frac{\partial }{\partial {\color{ForestGreen}{k}_{b}} } \sum_{ \vec{k}^{\prime} } {\color{red}w^{A}_{s,\vec k,\vec k'}} \left[\frac{\partial {\color{ForestGreen}f_0 (\vec k')}}{\partial {\color{ForestGreen}\vec k'}}\right]_c + {\color{ForestGreen}\tau^{s}_{\omega} \left( \tau^{s} \right)^{2}} \sum_{ \vec{k}^{\prime} } {\color{red}w^{A}_{s,\vec k,\vec k'}}  \left[\frac{\partial }{\partial {\color{ForestGreen}\vec k'}} \left(\frac{\partial {\color{ForestGreen}f_0 (\vec k')}}{\partial {\color{ForestGreen}\vec k'}}\right)_c \right]_b \bigg\}$     & ${\color{red}-}$          \\
Anomalous skew-scattering (ASN)   & $\frac{2 e^3\varepsilon_{adb}}{\hbar^2}\sum_{\vec k, \vec k', s} {\color{red}\Omega^{s}_{d}(\vec k)} {\color{ForestGreen}\tau_{s,\omega}^{2}} {\color{red}w^A_{s,\vec k,\vec k'}} \left[\frac{\partial {\color{ForestGreen}f_0 (\vec k')}}{\partial {\color{ForestGreen}\vec k'}}\right]_c$ & ${\color{ForestGreen}+}$   \\  
\hline \hline
\end{tabular}
\end{center}
\caption{Intraband nonlinear Hall responses and $PT$ symmetry. Green color indicates that the quantity is even under $PT$ symmetry, while red color means it is $PT$-odd. Similarly, ${\color{ForestGreen}+}$ indicates the response is even (i.e. allowed by symmetry), $-$ means it is odd (i.e. forbidden by symmetry).}
\label{Table:SM-PT-Classification}
\end{table}

\subsection{ASN in a 2D effective model of Dirac fermions in the antiferromagnetic metal CuMnAs}
\begin{figure}
\centering
\includegraphics[width=0.6\linewidth]{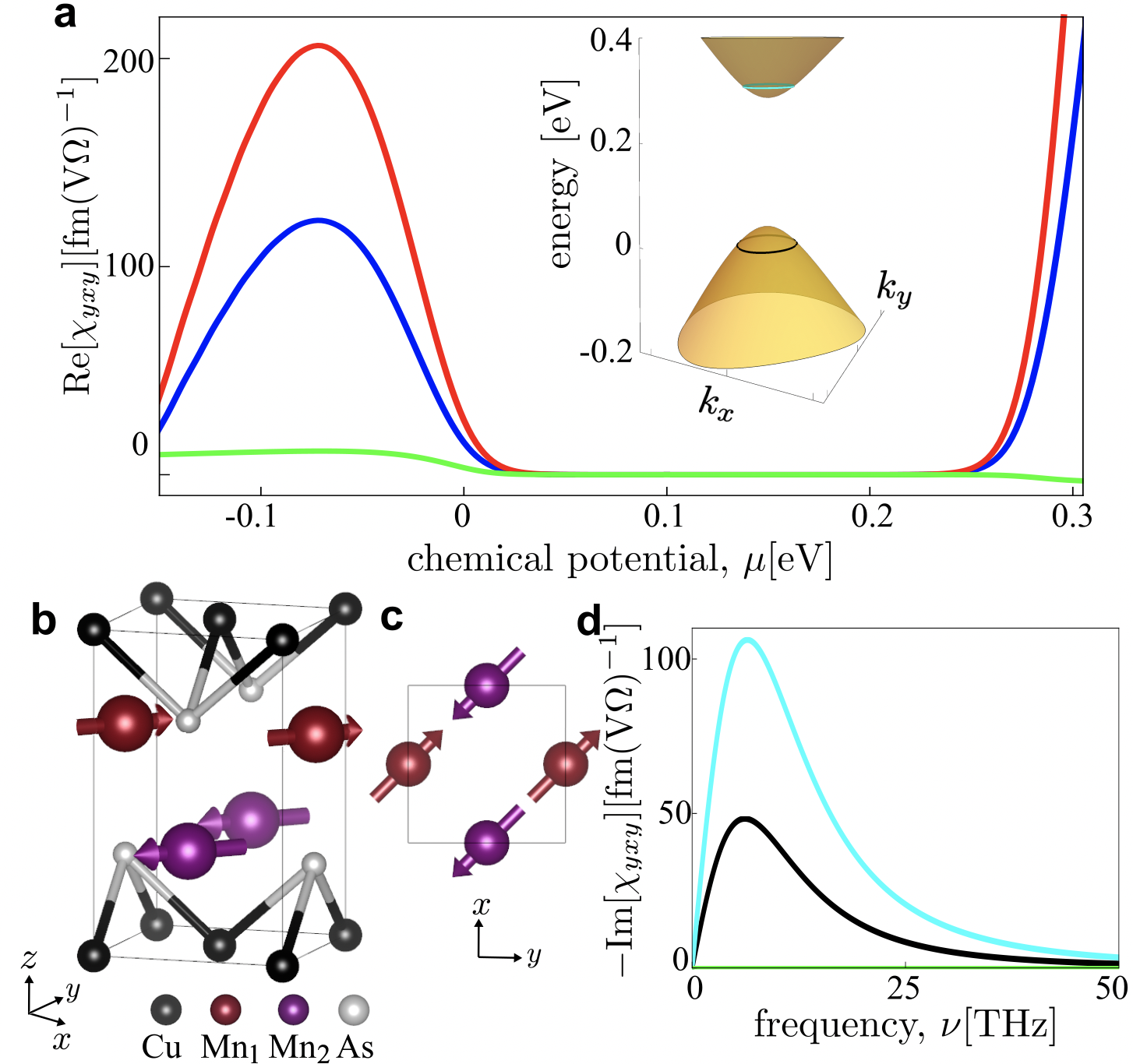}
\caption{{\bf ASN in 2D Dirac Fermions in the antiferromagnet CuMnAs.} (a) ASN and INH susceptibilities at varying chemical potential in the DC limit for Hamiltonian in Eq.~(\ref{eq:Jungwirth}). Blue curve for ASN with $V_{0} = 2.7 \times 10^{-13} \ \mathrm{cm}^{2} \mathrm{eV}$, red curve for ASN with  $V_{0} = 1.6 \times10^{-13} \ \mathrm{cm}^{2} \mathrm{eV}$, and green for INH. Due to mirror symmetry, $\chi_{xyx} = \chi_{xyy} = - \chi_{yxy} = - \chi_{yxx}$ for both ASN and INH. ASN dominates over INH. Note that here ASN and INH are of opposite signs in the conduction band. The relaxation time $\tau$ is around 15 fs for the blue curve and 40 fs for the red curve. Inset: Band structure of 2D CuMnAs with $\vec{n} \parallel [110]$. The material is gapped. (b) Lattice structure of 2D tetragonal CuMnAs with an antiferromagnetic order. (d) Imaginary part of ASN and INH susceptibilities. Cyan and black curves are for $\mu =-0.03$ and $0.3 \ \mathrm{eV}$, respectively, with the corresponding Fermi surfaces highlighted in the inset of (a). The imaginary part of ASN is nonzero and is sensitive to frequency, but INH (green curve) does not have an imaginary part.  Parameters: for (a) and (d), $n_{\mathrm{i}} = 4 \times 10^{9} \ \mathrm{cm}^{-2}$, lattice constant $a=0.38\ \mathrm{nm}$~\cite{Wadley2013}, $k_{\mathrm{B}}T=8\ \mathrm{meV}$; for (d), $V_{0} = 2.7 \times 10^{-13} \ \mathrm{cm}^{2} \mathrm{eV}$. }
\label{fig:CuMnAs}
\end{figure}

In this section we consider ASN in an effective model of Dirac fermions in a 2D antiferromagnetic metal tetragonal CuMnAs. In its antiferromagnetic phase, two Mn atoms connected by inversion symmetry have opposite orientation of magnetic moments, breaking {\it P} and {\it T} symmetry individually while preserving {\it PT} symmetry, see Fig.~\ref{fig:CuMnAs}b and c. Experiments have shown that the N\'{e}el vector describing the antiferromagnetic order can be reoriented with an electric current through spin orbit torque~\cite{Wadley2016,Godinho2018} and detected via second-order nonlinear responses~\cite{Godinho2018}. A single-layer quasi-2D tetragonal CuMnAs possesses Dirac fermions that can be modelled with the following tight-binding Hamiltonian~\cite{Jungwirth2017},
\begin{equation}
\label{eq:Jungwirth}
H(\vec{k}) = -2t \sigma_{x} \cos \frac{k_{x} a}{2} \cos \frac{k_{y} a}{2} - t' \left( \cos k_{x} a + \cos k_{y} a \right) +\lambda  \sigma_{z}  \left( s_{y} \sin k_{x} a - s_{x} \sin k_{y} a \right) + \sigma_{z} J_{n} \vec{s} \cdot \vec{n},
\end{equation}
where $t$ corresponds to nearest-neighbor hopping, $t'$ stands for next-nearest-neighbor hopping,  $\lambda$ is the next-nearest-neighbor spin-orbit coupling, $\vec{n}$ is the N\'{e}el vector, $a$ is the lattice constant, and Pauli matrices $\sigma$ and $s$ describe sublattices and spin degrees of freedom respectively~\cite{Jungwirth2017}. In this Hamiltonian, $P = \sigma_{x}s_{0}$, and $T = - i \sigma_{0}s_{y}K$. The last term in Eq.~(\ref{eq:Jungwirth}) breaks both {\it P} and {\it T}, but {\it PT} is respected. The parameters are estimated to be $\lambda = 0.8t$, $J_{n} = 0.6t$, and $t'=0.08t$, and $\vert {t} \vert = 1 \mathrm{eV}$~\cite{Jungwirth2017,Watanabe2021,Bhalla2022}. The band structure is sensitive to $\vec{n}$~\cite{Jungwirth2017}. Where as $\vec{n} \parallel [100]$ and $\vec{n} \parallel [010]$ result in two Dirac points, $\vec{n} \parallel [110]$ corresponds to a gapped phase, see Fig.~\ref{fig:CuMnAs}a (inset).

Unlike the model in the main text, real spin $s_{z}$ here is no longer a good quantum number. However, when the N\'{e}el vector lies in $x-y$ plane, we can still find decoupled {\it PT} partners with opposite pseudo-spins, where the pseudo-spin-z operator is $\sigma_{x}s_{z}$. Pseudo-spin block-diagonalizing the Hamiltonian, the semiclassical treatment remains applicable. The Hamiltonian in Eq.~(\ref{eq:Jungwirth}) can be expressed with five generators of the Clifford algebra $\Gamma_{1,2,3,4,5} = \{ \sigma_{x} s_{0}, \sigma_{y} s_{0}, \sigma_{z} s_{x}, \sigma_{z} s_{y}, \sigma_{z} s_{z} \}$ as
\begin{equation}
\label{eq:Methods-H}
H(\vec{k}) = -2t \cos \frac{k_{x} a}{2} \cos \frac{k_{y} a}{2}  \Gamma_{1} - t' \left( \cos k_{x} a + \cos k_{y} a \right) + \left( \lambda \sin k_{x} a + J_{n} n_{y} \right) \Gamma_{4} + \left(- \lambda  \sin k_{y} a + J_{n} n_{x} \right) \Gamma_{3}.
\end{equation}
Two consecutive unitary transformations, $U_{1} =  \frac{1}{\sqrt{2} } \left( \sigma_{0}s_{0} + \Gamma_{2} \Gamma_{4} \right) $ and  $U_{2} =\frac{1}{\sqrt{2} } \left( \sigma_{0} s_{0} + i \sigma_{0} s_{y} \right)$ help to block-diagonalize the Hamiltonian above,
\begin{equation}
\label{eq:rotated}
H'(\vec{k}) = U_{2} U_{1} H(\vec{k}) U_{1}^{\dagger} U_{2}^{\dagger} = -2t \cos \frac{k_{x} a }{2} \cos \frac{k_{y} a}{2}  \Gamma_{1} - t' \left( \cos k_{x} a+ \cos k_{y} a \right) + \left( \lambda \sin k_{x} a+ J_{n} n_{y} \right) \Gamma_{2} + \left(- \lambda  \sin k_{y} a+ J_{n} n_{x} \right) \Gamma_{5},
\end{equation}
where $s_{z}$ is now a good quantum number. Reversing the unitary transformations, we see that the corresponding pseudo-spin-z operator in Eq.~(\ref{eq:Methods-H}) is $-\sigma_{x}s_{z}$.

Fig.~\ref{fig:CuMnAs}a shows the nonlinear susceptibilities of the Hamiltonian in Eq.~(\ref{eq:Jungwirth}) with $\vec{n} \parallel [110]$ at the DC limit. Due to the mirror symmetry $\left( k_{x}, k_{y} \right) \to \left( -k_{y}, -k_{x} \right)$, we only have one independent nonlinear susceptibility for ASN and INH mechanisms, $\chi_{xyx} = \chi_{xyy} = - \chi_{yxy} = - \chi_{yxx}$. Even with the smaller $\tau$ here, the magnitude of ASN susceptibility wins over that of the INH. As shown in Fig.~\ref{fig:CuMnAs}a, both ASN and INH contribution grows as the chemical potential $\mu$ moves from edge into the band. In Fig.~\ref{fig:CuMnAs}d, the imaginary part of ASN susceptibility is nonzero and is sensitive to frequency, which can help to distinguish ASN from INH.

\end{document}